\documentclass[%
 reprint,
superscriptaddress,
 amsmath,amssymb,
 aps,
]{revtex4-2}

\usepackage{graphicx}
\usepackage{dcolumn}
\usepackage{bm}
\usepackage{gensymb}
\usepackage[colorlinks=true, allcolors=blue]{hyperref}
\usepackage[all]{hypcap}       

\usepackage{microtype}

\usepackage{booktabs}
\usepackage{makecell}
\usepackage{hyperref}
\usepackage{xr-hyper}

\usepackage[utf8]{inputenc}
\usepackage[T1]{fontenc}
\usepackage{mathptmx}
\usepackage{etoolbox}
\usepackage{upgreek}
\usepackage[version = 4]{mhchem}

\usepackage{xcolor}
\usepackage{graphicx}
\usepackage{dcolumn}
\usepackage{bm}
\usepackage{upgreek}
\usepackage{booktabs}
\usepackage{makecell}  
\usepackage[version = 4]{mhchem}

\renewcommand{\textcolor}[2]{#2}
\begin{document}

\preprint{APS/123-QED}

\title{Nitride crystalline barrier epitaxial SIS Josephson junctions by MBE%
}


\author{\underline{Manas Verma}}

\email{mv388@cornell.edu}

\affiliation{School of Applied and Engineering Physics, Cornell University, Ithaca, NY 14853, USA}
\thanks{Underlined authors contributed equally.} 

\author{\underline{Anand Ithepalli}}

\affiliation{Department of Materials Science and Engineering, Cornell University, Ithaca, NY 14853, USA}
\affiliation{Kavli Institute at Cornell for Nanoscale Science, Cornell University, Ithaca, NY 14853, USA}

\author{\underline{Amit Rohan Rajapurohita}}

\affiliation{School of Applied and Engineering Physics, Cornell University, Ithaca, NY 14853, USA}
 
\author{Eegene Clara Chung}
\affiliation{Department of Physics, Cornell University, Ithaca, NY 14853, USA}

\author{David A. Garcia-Wetten}
\affiliation{Department of Applied Physics, Stanford University, Stanford, California 94305, USA}

\author{Naomi Pieczulewski}
\affiliation{Department of Materials Science and Engineering, Cornell University, Ithaca, NY 14853, USA}

\author{Danqing Wang}
\affiliation{Department of Electrical Engineering, Yale University, New Haven, Connecticut 06511, USA}

\author{John G. Wright}
\affiliation{Department of Materials Science and Engineering, Cornell University, Ithaca, NY 14853, USA}

\author{Archit Shah}
\affiliation{Electrical and Computer Engineering Department, Auburn University, Auburn, AL 36849 USA}

\author{Jimy Encomendero}
\affiliation{School of Electrical and Computer Engineering, Cornell University, Ithaca, NY 14853, USA}

\author{Rishabh Singh}
\affiliation{School of Applied and Engineering Physics, Cornell University, Ithaca, NY 14853, USA}

\author{Chandrashekar Savant}
\affiliation{Department of Materials Science and Engineering, Cornell University, Ithaca, NY 14853, USA}

\author{Hangwang Yang}
\affiliation{Department of Materials Science and Engineering, Cornell University, Ithaca, NY 14853, USA}

\author{Anthony Hyatt}
\affiliation{Department of Materials Science and Engineering, Cornell University, Ithaca, NY 14853, USA}

\author{Jaehong Choi}
\affiliation{School of Applied and Engineering Physics, Cornell University, Ithaca, NY 14853, USA}

\author{Thomas Varghese}

\affiliation{Department of Materials Science and Engineering, Cornell University, Ithaca, NY 14853, USA}

\author{Kushagra Aggarwal}
\affiliation{School of Applied and Engineering Physics, Cornell University, Ithaca, NY 14853, USA}

\author{Paul B. Welander}
\affiliation{SLAC National Accelerator Laboratory, Menlo Park, California 94025, USA}

\author{Apurva Mehta}
\affiliation{Linac Coherent Light Source, SLAC National Accelerator Laboratory, Menlo Park, California 94025, USA}

\author{David Schuster}
\affiliation{Department of Applied Physics, Stanford University, Stanford, California 94305, USA}

\author{Valla Fatemi}
\affiliation{School of Applied and Engineering Physics, Cornell University, Ithaca, NY 14853, USA}
\affiliation{Kavli Institute at Cornell for Nanoscale Science, Cornell University, Ithaca, NY 14853, USA}

\author{Gregory D. Fuchs}
\affiliation{School of Applied and Engineering Physics, Cornell University, Ithaca, NY 14853, USA}
\affiliation{Kavli Institute at Cornell for Nanoscale Science, Cornell University, Ithaca, NY 14853, USA}

\author{Michael C. Hamilton}
\affiliation{Electrical and Computer Engineering Department, Auburn University, Auburn, AL 36849 USA}

\author{Hong X. Tang}
\affiliation{Department of Electrical Engineering, Yale University, New Haven, Connecticut 06511, USA}

\author{David A. Muller}

\affiliation{School of Applied and Engineering Physics, Cornell University, Ithaca, NY 14853, USA}
\affiliation{Kavli Institute at Cornell for Nanoscale Science, Cornell University, Ithaca, NY 14853, USA}

\author{Huili Grace Xing}
\affiliation{Department of Materials Science and Engineering, Cornell University, Ithaca, NY 14853, USA}
\affiliation{Kavli Institute at Cornell for Nanoscale Science, Cornell University, Ithaca, NY 14853, USA}
\affiliation{School of Electrical and Computer Engineering, Cornell University, Ithaca, NY 14853, USA}

\author{Debdeep Jena}

\affiliation{Department of Materials Science and Engineering, Cornell University, Ithaca, NY 14853, USA}
\affiliation{Kavli Institute at Cornell for Nanoscale Science, Cornell University, Ithaca, NY 14853, USA}
\affiliation{School of Electrical and Computer Engineering, Cornell University, Ithaca, NY 14853, USA}

\date{\today}

\begin{abstract}


\textcolor{blue}{Superconductor-insulator-superconductor Josephson junctions with crystalline tunnel barriers have remained elusive by the epitaxial techniques such as MOCVD or MBE that produce the junction materials and devices powering today's electronic and photonic systems like transistors and lasers. Here, we overcome the key materials hurdle to realize the first functional nitride crystalline barrier Josephson junctions by MBE.  Despite an AlN barrier thickness $\sim $4.5 nm, the NbN/AlN/NbN trilayers exhibit coherent Josephson tunneling with critical current densities of $\sim$17 A cm$^{-2}$, an $I_\mathrm{c}R_\mathrm{n}$ product of $\sim$2.5 mV, and a superconducting gap voltage of $\sim$3.7 mV with a BCS-like gap evolution with temperature. Hysteretic current–voltage characteristics and Fraunhofer current modulation with magnetic-field establish underdamped SIS Josephson behavior.  The realization of crystalline barrier Josephson junctions by MBE opens the doors to a richer materials pallette with alloys and superlattice electrodes and tunnel barriers, as well as potential monolithic integration of superconducting quantum circuits with active electronic and photonic devices such as HEMTs and Lasers.}

\end{abstract}

\maketitle


\textcolor{blue}{Since Josephson's proposal in 1962 \cite{Josephson1962}, and experimental realization the year after \cite{Anderson1963}, Josephson junctions (JJs) have evolved to become central components for quantum-limited detectors and mixers \cite{Fagaly2006, Zmuidzinas2004, Tucker1985}, high-frequency superconducting digital electronics \cite{Likharev2012}, and superconducting quantum circuits \cite{Clarke2008, Devoret2013, Koch2007}.  Among the many material variants, superconductor-insulator-superconductor (SIS) JJs exhibit the macroscopic quantum properties desired for contemporary superconducting qubits \cite{Martinis1985, Devoret1985}. Today's superconducting qubits use JJs with aluminum electrodes separated by an amorphous $\mathrm{AlO}_{\mathrm{x}}$ barrier \cite{Bland2025, Koch2007, Kjaergaard2020}.}  

\textcolor{blue}{To realize crystalline tunnel barrier JJs desired for surmounting quantum coherence limits of amorphous barriers, sputtering and ALD have been used \cite{Wang2013, DanqingWang2026, Wilt2017}.  Metal-organic chemical vapour deposition (MOCVD) and molecular beam epitaxy (MBE) are the only techniques today that produce materials with the stringent requirements needed for active gain--for example--photonic gain in semiconductor lasers and electronic gain in high electron-mobility transistors (HEMTs). Successful realization of JJs by these techniques will open the gates to vast untapped materials combinations, ranging from monolithic super-semi junctions for gatemons, combining JJs with HEMTs, light-emitting diodes (LEDs), and lasers, and unchartered materials combinations for the JJ electrode and barrier materials such as superlattices, quantum dots, and van der Waals tunnel barriers.}


\begin{figure*}[htbp]
    \centering
    \includegraphics[width=\textwidth]{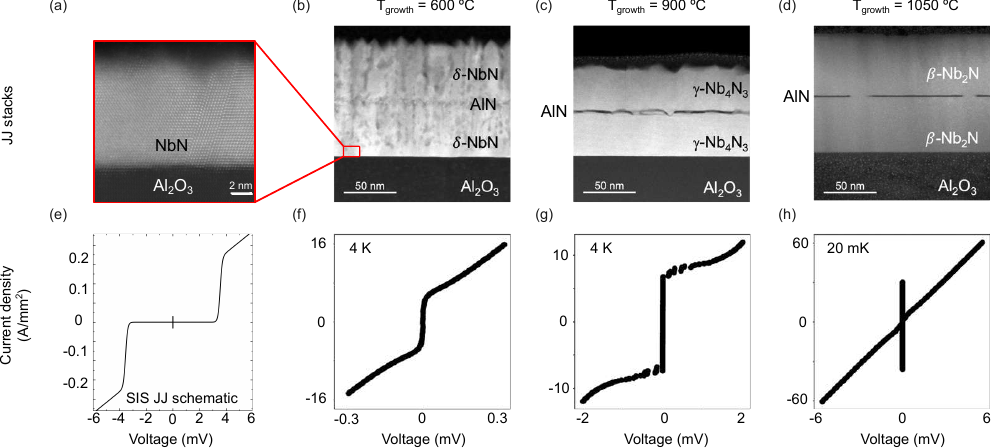}
\caption{\label{fig:Failed_JJ}
\textbf{Previous Josephson-junction growth attempts: Growth-temperature effects on NbN/AlN/NbN structure and transport.} 
(a) Magnified cross-sectional STEM image of the NbN/substrate interface from panel (b), showing approximately the first 10~nm of highly coherent NbN growth without apparent columnar features. 
(b--d) Cross-sectional STEM images of NbN/AlN/NbN heterostructures grown at 600~$^{\circ}$C, 900~$^{\circ}$C, and 1050~$^{\circ}$C, respectively, showing the evolution of the NbN electrodes from cubic $\delta$-NbN to tetragonal $\gamma$-Nb$_4$N$_3$ and hexagonal $\beta$-Nb$_2$N. Despite the different NbN structural phases, the AlN barrier remains highly inhomogeneous, exhibiting pronounced thickness variations, local thinning, and  structural shorts. 
(e) Schematic of an ideal voltage-biased Josephson-junction $I$--$V$ characteristic illustrating the hysteretic response expected for an underdamped junction. 
(f--h) Corresponding experimental current-biased $I$--$V$ characteristics of the heterostructures shown in panels (b--d), exhibiting non-SIS transport consistent with the structural imperfections of the AlN barrier.}
\end{figure*}

\textcolor{blue}{The first hurdle, the realization of MOCVD- or MBE-grown Josephson junctions simultaneously exhibiting three key signatures of underdamped behavior required for qubits— SIS tunnelling characteristics, hysteretic current–voltage behavior, and Fraunhofer diffraction of the supercurrent—has remained elusive, despite considerable efforts \cite{Supple2024, Lachowski2025, Knudsen2026GeJJ, Li2017MgB2}.  In this work, we report the first realization of crystalline tunnel-barrier NbN/AlN/NbN Josephson junctions by MBE that pass all three of these stringent tests. This was made possible by identifying, through electron microscopy, the origin of defects
that prevented earlier efforts by various groups, including ours, from achieving SIS JJ behavior. We manage defect formation
by staying below the critical thicknesses of the crystalline superconducting layers and barriers. To highlight the impact of crystalline barriers and MBE, we further demonstrate that the tunneling current scales with junction area and varies by over
two orders of magnitude with crystalline tunnel-barrier thickness.

While MOCVD waits for the development of appropriate precursors, MBE has been used to explore control over layer thickness, crystallographic orientation, interface structure and barrier composition during the growth of superconductor layers and junctions \cite{Lachowski2025, Terai2026, Oh2006, Nakamura2011, Wang2013, Kim2021, Supple2024}.  MBE has historically enabled landmark advances in materials science and device applications, including magnetic multilayers exhibiting giant magnetoresistance (GMR) \cite{Fert2008}, and the fractional quantum Hall effect in high mobility semiconductor quantum wells \cite{Stormer1999, Tsui1999, Tsui1982}.  Its uses in realizing epitaxial superconductor–semiconductor hybrids indicate the potential for all-epitaxial Josephson junctions and integrating them with the broader nitride materials platform \cite{Krogstrup2015, Jena2019, Lachowski2025}.}

\textcolor{blue}{NbN is a particularly attractive superconductor for JJ electrodes because of its high critical temperature, large superconducting gap, and chemical compatibility and near lattice-matching with insulating and semiconducting nitrides such as AlN \cite{Jena2019, Kim2021, Wright2021, Terai2026, Wang2013}.  Yet basic NbN/AlN/NbN JJ properties realized by DC magnetron sputtering remain elusive by MBE \cite{Terai2026, Kim2021, Nakamura2011, Wang2013}.}

\textcolor{blue}{Realizing an AlN tunnel barrier on NbN by MBE presents a critical materials challenge.  Under practical MBE conditions, the NbN layer serving as the bottom JJ electrode was recently found to develop a columnar microstructure with increasing surface roughness as the film thickens \cite{Ithepalli2026, Supple2024}.  Because the tunneling current depends exponentially on the local barrier thickness, small variations in the morphology of the underlying NbN electrode produces local thinning or discontinuities in the AlN barrier layer, leading to electrical shorts \cite{Simmons1963, Supple2024}.  In addition, Al-rich growth conditions are unsuitable for AlN deposition directly on NbN because Al chemically reacts and deteriorates the NbN surface, necessitating nitrogen-rich growth conditions \cite{Pogrebnjak2017, Lucadamo2001}.  Such N-rich conditions tend to promote three-dimensional island growth of AlN in plasma--assisted MBE, resulting in a rough and nonuniform upper NbN/AlN interface; this makes the formation of a continuous tunnel barrier particularly challenging \cite{Koblmuller2003}.  These combined morphological and chemical constraints highlight why prior demonstrations of SIS behavior in NbN/AlN/NbN Josephson junctions grown entirely by MBE have not succeeded \cite{Lachowski2025, Supple2024}.}

\begin{figure*}
\centering
\includegraphics[width=\textwidth]{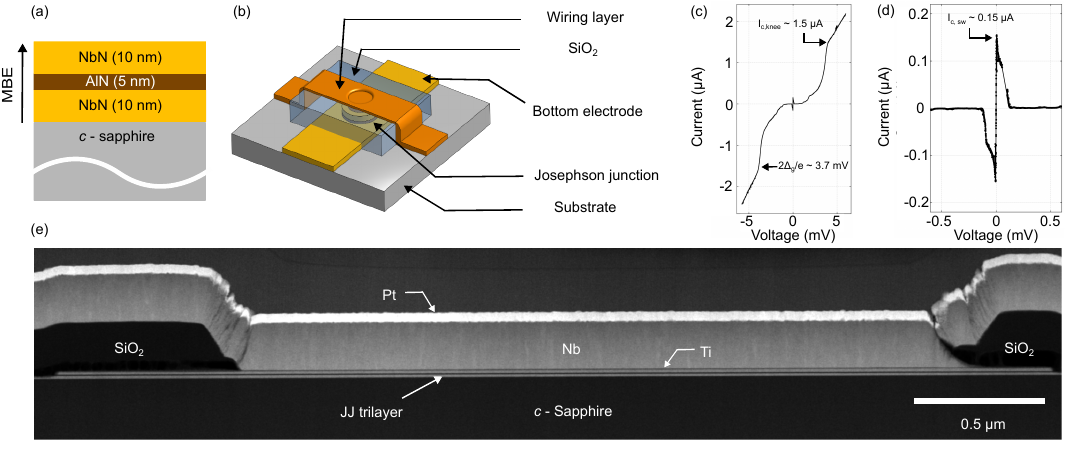}
\caption{\label{fig:overview}
\textbf{Epitaxial NbN/AlN/NbN Josephson junction platform.}
\textbf{(a)} Schematic of the plasma-assisted molecular beam epitaxy-grown trilayer heterostructure, consisting of a bottom NbN electrode, an AlN tunnel barrier and a top NbN electrode on sapphire.
\textbf{(b)} Schematic of the completed Josephson junction device after mesa definition, contact formation and wiring.
\textbf{(c)} Voltage-biased current--voltage characteristic measured at 18~mK, showing clear SIS tunneling behavior with a quasiparticle gap voltage of approximately 3.7~mV.
\textbf{(d)} Magnified low-voltage \(I\)–\(V\) characteristics showing a switching current of \(I_{\mathrm{sw}} \approx 0.15\,\mu\mathrm{A}\).
\textbf{(e)} Cross-sectional STEM image of the fabricated junction, confirming a continuous AlN barrier separating the NbN electrodes and the absence of electrically shorted regions across the junction area.
Together, these results establish an all-epitaxial nitride Josephson junction stack with a structurally continuous tunnel barrier and robust SIS transport.
}
\end{figure*}

\begin{figure*}[htbp]
    \centering
    \includegraphics[width=\textwidth]{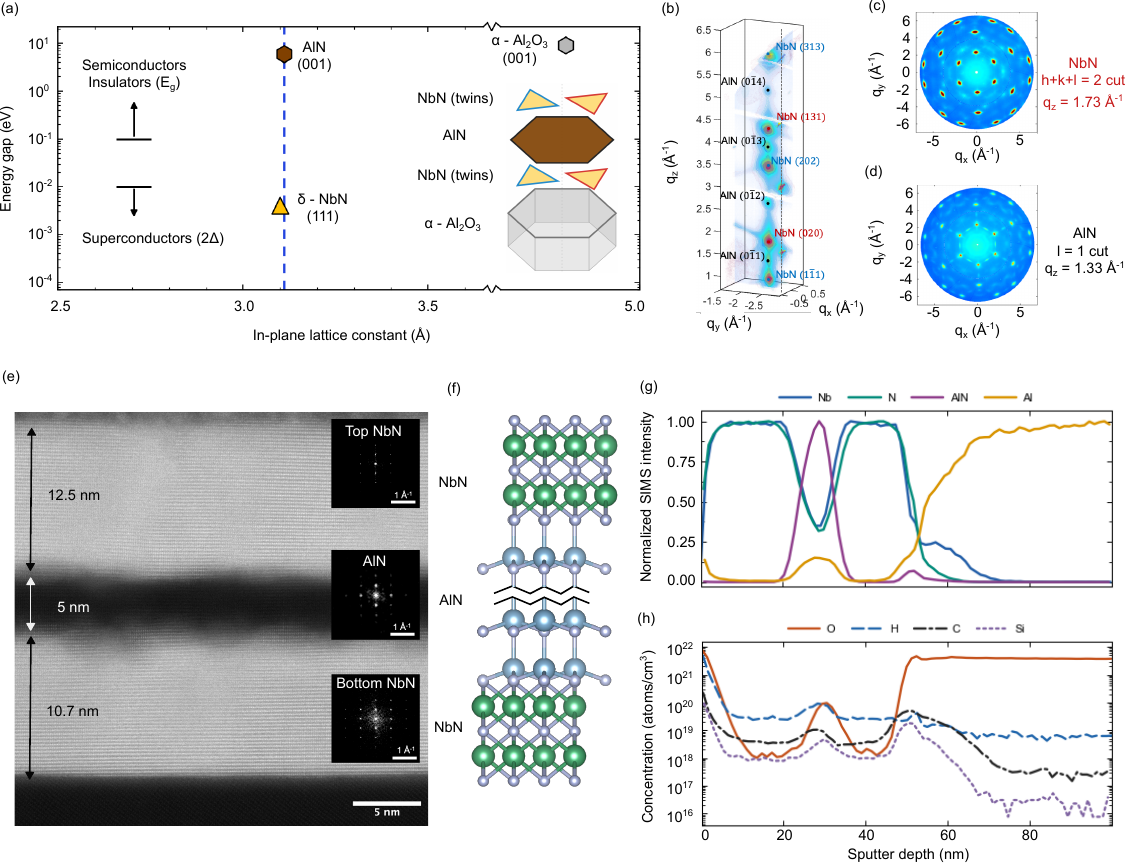}
\caption{\label{fig:TEM}\textbf{Materials selection and structural characterization of the epitaxial NbN/AlN/NbN trilayer.}
\textbf{(a)} Energy gap versus in-plane lattice constants for the superconductors, semiconductors and insulators of this work. The near lattice match between NbN(111) and wurtzite AlN ($\sim0.3\%$ mismatch) favours epitaxial NbN/AlN interfaces, whereas NbN relaxes with respect to sapphire. Inset: Schematic of the epitaxial registry and 180° twinning of NbN on wurtzite AlN and sapphire.
\textbf{(b)} Off-specular crystal truncation rod measured by high intensity X-ray shows the epitaxial registry of the NbN and AlN layers. Two 180°-twinned NbN domains are indicated in blue and red. The films are relaxed with respect to the substrate, as evidenced by the in-plane offset of the substrate truncation rod (grey dashed line). AlN reflections are indexed according to the wurtzite structure. \textbf{(c, d)} The agreement between the observed AlN reflections and the expected reciprocal lattice positions confirms the wurtzite crystal structure of the AlN layer. Lateral cuts of the reciprocal space map showing NbN and AlN reflections, respectively. The oblong features between AlN reflections are rods connecting NbN reflections, rather than discrete Bragg reflections satisfying the Laue conditions.
\textbf{(e)} Cross-sectional STEM image showing a continuous AlN barrier between the NbN electrodes. FFTs in the inset show distinct diffraction peaks indicative of crystalline layers.
\textbf{(f)} VESTA atomic model of the trilayer stacking sequence showing an Al-polar AlN barrier region.
\textbf{(g)} Normalized SIMS depth profiles of Nb, N , Al and AlN.
\textbf{(h)} SIMS concentration profiles of O, H, C and Si. Interfacial smearing is due to SIMS depth resolution. The elemental distribution is consistent with the expected NbN/AlN/NbN trilayer on $\mathrm{Al_2O_3}$.}
\end{figure*}

\textcolor{blue}{Fig.\ref{fig:Failed_JJ} summarizes our own unsuccessful attempts to realize MBE NbN/AlN/NbN JJs highlighting these challenges.  The top panel indicates the JJ trilayer stacks, and the bottom panel shows the desired SIS JJ characteristic on the left, and the three measured characteristics.  Figures \ref{fig:Failed_JJ}(b–d) show cross-sectional scanning transmission electron microscopy (STEM) images of film stacks grown at increasing temperature, resulting in distinct NbN structural phases from cubic $\delta$-NbN at 600~$^{\circ}$C to tetragonal $\gamma$-Nb$_4$N$_3$ at 900~$^{\circ}$C to hexagonal $\beta$-Nb$_2$N at 1050~$^{\circ}$C. The AlN barrier remains discontinuous in all cases, exhibiting pronounced thickness variations, local thinning, and shorts in the STEM images.  Each of the three corresponding $I-V$ characteristics in Figs. \ref{fig:Failed_JJ}(f–h) display non-SIS transport behavior.  Figure \ref{fig:Failed_JJ}(e) schematically shows the desired voltage-biased $I$--$V$ characteristics of a JJ exhibiting a supercurrent branch at zero voltage, followed by an insulating voltage, and then normal state resistance \cite{Tinkham1996}.}

\textcolor{blue}{Yet, hidden in the STEM images, is a potential solution.  Fig. \ref{fig:Failed_JJ}(a) shows a magnified view of the metal--substrate interface in Fig. \ref{fig:Failed_JJ}(b).  We discovered that the first $\sim$10 nm of NbN adjacent to the substrate exhibits highly coherent growth with no apparent columnar features \cite{Ithepalli2026}.  Can restricting the NbN to below this thickness enable the first MBE JJs?  Here we show that indeed this path leads to the first fully functional crystalline barrier trilayer NbN/AlN/NbN JJs by MBE.}

\textcolor{blue}{Figure \ref{fig:overview} shows how we constructed the all-epitaxial crystalline barrier NbN/AlN/NbN trilayer JJs by MBE in this work and its DC $I-V$ characteristics.  Using plasma-MBE we grew the NbN/AlN/NbN trilayer heterostructure shown in Figure \ref{fig:overview} (a) on c-plane sapphire.  Because superconducting properties of NbN depend on its structural phase, we obtained cubic $\delta$-NbN phase due to its higher $T_{\rm c}$ [Fig. S3 (a)] by using MBE growth conditions we have reported earlier to obtain phase pure structural variants \cite{Wright2021, Wright2023}.  We grew the entire trilayer at a constant substrate temperature of $600^{\circ}\mathrm{C}$ to stabilize $\delta$-NbN.  We restricted the bottom electrode thickness to $<$12 nm to suppress formation of columnar grains and provide a smooth surface for the ultrathin AlN tunnel barrier \cite{Ithepalli2026}.  We then deposited a 5-nm-thick AlN barrier under N-rich conditions (Al:N $\sim$ 1:4).  We chose this growth condition to balance competing interface constraints: excess Al induces Nb–Al intermetallic formation \cite{Lucadamo2001, Pogrebnjak2017}, whereas N-rich condition promotes three-dimensional AlN island growth and barrier roughness \cite{Koblmuller2003}.  Optimal N-rich growth conditions minimize interfacial reaction and leverage the close lattice matching between NbN and AlN to promote island coalescence into a continuous tunnel barrier (Fig. \ref{fig:TEM} (a)).  We then deposited a 10-nm-thick NbN counter-electrode using the same conditions as the bottom NbN electrode, completing the all-\textit{in situ} NbN/AlN/NbN junction stack of Figure \ref{fig:overview} (a).}

\textcolor{blue}{We constructed the JJ device structure shown in Figure \ref{fig:overview} (b) from the MBE trilayer using fabrication steps described in \cite{Choi2025}.  The figure shows the mesa-defined circular regions that form the JJs and independent electrical contacts that provide access to the top and bottom electrodes.  We used a SiO$_2$ scaffold to electrically isolate the wiring layer from the exposed bottom electrode and prevent shorting outside the junction area.  The cross-sectional STEM image of the JJ in Figure \ref{fig:overview} (e) shows a continuous and uniform NbN / AlN / NbN junction trilayer, and the Ti (3 nm) / Nb (200 nm) / Ti (3 nm) / Pt (30 nm) wiring stack.  The AlN tunnel barrier remains smooth and continuous across the junction area, with no discernible discontinuities, pinholes, or shorts between the NbN electrodes.}

\textcolor{blue}{In Figures~\ref{fig:overview} (c) and (d) we show the dc $I-V$ characteristics of a 3-$\mu$m-diameter JJ that we measured with voltage bias at 18 mK.  We finally obtain the JJ characteristics that was sketched in Figure \ref{fig:Failed_JJ} (e).  We observe the characteristic strongly nonlinear behavior desired in a SIS junction, with suppressed current in the subgap regime followed by a pronounced increase in quasiparticle current near the sum-gap voltage.  We observe the gap edge at $V_{\rm gap}=3.7$ mV, which corresponds to a superconducting gap $\Delta \approx 1.85$ meV per NbN electrode, beyond which the current transitions towards the normal quasiparticle branch.  Using the measured $T_{\mathrm{c}}=10.82$~K (see Fig. S3 (a)) gives $2 \Delta / k_{\rm B}T_{\rm c} \approx 3.96$, consistent with the enhanced coupling previously observed in NbN \cite{Kihlstrom1985}.}

\textcolor{blue}{While the subgap current in Figure~\ref{fig:overview} (d) is small at low voltages, it becomes substantial for higher voltages near the sum-gap as seen in Figure~\ref{fig:overview} (c).  We observe this despite a structural thickness of $\sim$5~nm of the AlN barrier, independently confirmed by X-ray shown in Fig. \ref{fig:TEM} (b, c, d) and cross-sectional STEM in Fig. \ref{fig:TEM} (e).  This portion of the current is reminiscent of transparent, thin-barrier junctions, suggesting that the effective electrical tunneling barrier differs from its geometric thickness \cite{Kleinsasser1993, Kleinsasser1994}. The electronic polarization of wurtzite AlN can substantially alter the electronic tunneling thickness compared to the metallurgical one to cause enhanced transparency \cite{Simon2009}, but at this stage this is our (yet to be proven) hypothesis.}

\textcolor{blue}{The magnified view in Figure~\ref{fig:overview}(d) of the low-voltage Josephson response shows a switching current of $I_{\mathrm{sw}}\approx0.15~\mu\mathrm{A}$.  Using the measured superconducting gap $\Delta=$1.85 meV and normal-state resistance $R_{\mathrm{N}}=2.58$ k$\Omega$,  the intrinsic critical current from the Ambegaokar--Baratoff relation \cite{Ambegaokar1963} is 
\begin{equation}
I_\mathrm{c} = \frac{\pi \Delta}{2eR_\mathrm{N}} \approx 1.12~\mu\mathrm{A}.
\label{eq:ambegaokar_baratoff}
\end{equation}
The experimentally observed switching current is nearly an order of magnitude smaller, consistent with premature switching driven by phase fluctuations and noise in the electromagnetic environment of the junction \cite{Fulton1974, Grabert1998, Kautz1990}.}

\textcolor{blue}{To understand the transport properties of the NbN/AlN/NbN Josephson junction, we studied its crystallographic and chemical properties.  Figure~\ref{fig:TEM} (a) plots the energy gap against in-plane lattice constants of the epilayer materials, using a broken axis to include sapphire. The three-fold basal-plane symmetry of \textit{c}-plane sapphire favours [111]-oriented cubic NbN indicated by the triangular symbols, to which a [0001] oriented AlN tunnel barrier is nearly lattice matched and of hexagonal symmetry. The large in-plane lattice mismatch of $\sim13\%$ of both with sapphire results in strain relaxation of the bottom NbN electrode. Because the lattice constant of NbN(111) $a_{\mathrm{NbN}}/\sqrt{2}\approx3.10~\text{\AA}$ is nearly commensurate with wurtzite AlN  $a_{\mathrm{AlN}}\approx3.11~\text{\AA}$, a mismatch of only $\sim0.3\%$ between them epitaxially stabilizes the AlN barrier.}

\textcolor{blue}{Were these epitaxial relations actually obtained? We performed high intensity X-ray reciprocal space mapping (RSM) to investigate. Fig. \ref{fig:TEM} (b) shows an off-specular section of the RSM demonstrating the relaxation of the NbN/AlN/NbN heterostructure film with the sapphire substrate.  We find that the AlN and NbN truncation rods are matched in-plane. Figs. \ref{fig:TEM} (c, d) show in-plane sections of the RSM which confirms the wurtzite crystal structure of the AlN film, proving that the NbN/AlN/NbN trilayer is indeed a crystalline barrier JJ.  Fig. S1 (c) shows the indexing of the wurtzite AlN phase across the entire RSM.} 

\begin{figure}[htbp]
    \includegraphics[width=\columnwidth]{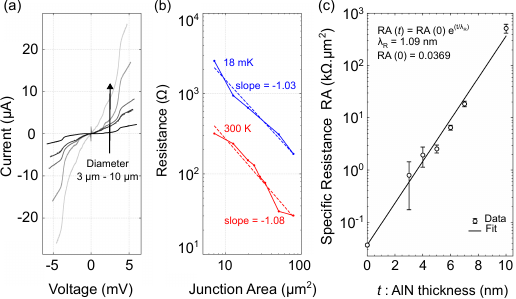}
    \caption{\label{fig:area_scaling}
    \textbf{Area and tunnel barrier scaling of MBE NbN/AlN/NbN JJs.}
    \textbf{(a)} Voltage-biased current--voltage characteristics measured at 18~mK for junctions with different device areas. The normal-state resistance decreases systematically with increasing junction area, demonstrating control of the junction current through lithographic variation of junction size.
    \textbf{(b)} Junction normal-state resistance, measured above the superconducting gap voltage ($V_{\mathrm{gap}} = 3.7$~mV), as a function of device area at room temperature and low temperature.
    \textbf{(c)} Resistance--area product ($RA$) as a function of AlN barrier thickness. The solid line represents an exponential fit, yielding $2\kappa = 0.921~\mathrm{nm}^{-1}$ and an extrapolated intercept $RA(0) = 0.0369~\mathrm{k\Omega\,\mu m^2}$.
    }
\end{figure}

\textcolor{blue}{We obtained further confirmation of the crystallinity of the trilayer JJ with cross-sectional STEM shown in Fig. \ref{fig:TEM} (e), which shows a continuous AlN barrier separating the top and bottom NbN electrodes.  Fast-Fourier transforms (FFTs) performed on the signals from the individual layers shown in the inset of Fig. \ref{fig:TEM} (e) confirms the crystalline nature of the heterostructure.  The top and bottom NbN electrodes exhibit distinct diffraction reflections, while the AlN barrier displays a well-defined array of discrete spots confirming the barrier is crystalline. The AlN FFT is consistent with observation along the $[1\bar{1}00]$ zone axis of the hexagonal lattice.  For wurtzite AlN, with lattice parameters $a \approx 3.11~\text{\AA}$ and $c \approx 4.98~\text{\AA}$, the expected interplanar spacings in this projection are $d_{0002} = c/2 \approx 2.49~\text{\AA}$, and $d_{11\bar{2}0} = a/2 \approx 1.56~\text{\AA}$, corresponding to a reciprocal-spacing ratio of $ q_{11\bar{2}0} / q_{0002} \approx 1.60$, consistent with the anisotropy observed in the FFT. These demonstrate the crystalline order across the NbN/AlN/NbN heterostructure. Fig. \ref{fig:TEM} (f) shows a corresponding VESTA atomic-model showing the stacking sequence across the heterostructure.}

\begin{figure}[htbp]
    \centering
    \includegraphics[width=\columnwidth]{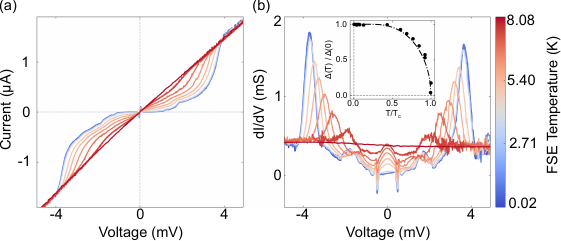}
\caption{\label{fig:gap_temperature}
\textbf{T-dependent current-voltage characteristics, numerically extracted differential conductance and superconducting gap.}
\textbf{(a)} Temperature-dependent voltage-biased current--voltage characteristics, showing the progressive suppression and closing of the quasiparticle gap with increasing temperature. The disappearance of the gap at elevated temperature confirms its superconducting origin.
\textbf{(b)} Differential conductance spectra numerically extracted as a function of temperature, showing the corresponding closing of the superconducting gap and the evolution of the quasiparticle tunnelling features.
The inset shows the extracted superconducting gap as a function of temperature, compared with the Bardeen--Cooper--Schrieffer temperature dependence. The BCS curve is calculated independently and is not fitted to the data, demonstrating excellent agreement between the measured gap evolution and conventional superconducting theory.
}
\end{figure}

\begin{figure*}[htbp]
    \centering
    \includegraphics[width=\textwidth]{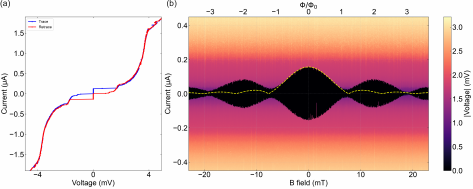}  
\caption{\label{fig:gap_temperature_fraunhofer}
\textbf{Current-biased current-voltage characteristics and magnetic-field modulation of epitaxial NbN/AlN/NbN Josephson junctions.}
\textbf{(a)} Current-biased current-voltage characteristics, showing hysteretic behavior at 0 T B-field. The highest critical current extracted from trace and re-trace of this experimental curve is used for observing Fraunhofer pattern at different magnetic fields.
\textbf{(b)} Magnetic-field dependence of the critical current, showing a Fraunhofer diffraction pattern. The field modulation of the critical current branch confirms coherent Josephson tunnelling through the junction barrier.
}
\end{figure*}

\textcolor{blue}{Figs. \ref{fig:TEM} (g, h) show the secondary-ion mass spectrometry (SIMS) depth profiles of the NbN/AlN/NbN trilayer and the underlying sapphire substrate. The Nb and N signals remain high within the NbN electrodes, while the Al signal increases across the AlN barrier and subsequently dominates in the sapphire substrate. The apparent broadening of the interfaces is attributed to the finite depth resolution and ion-beam-induced mixing inherent to the SIMS measurement. Fig. \ref{fig:TEM} (g) also reveals a pronounced AlN-related signal within the tunnel barrier, attributed to the sputtering and detection of AlN as a molecular species. Furthermore, Fig.~\ref{fig:TEM} (h) shows peaks in the hydrogen and oxygen concentrations within the AlN barrier, reaching approximately $10^{20}\,\mathrm{cm^{-3}}$.  Enhanced impurity concentrations are observed both within the AlN barrier and near the NbN--sapphire interface, suggesting localized impurity accumulation at these regions. In the supplementary information we discuss the possible implications of these impurity distributions for the electronic transport properties of the JJs.  Nevertheless, the overall elemental distribution---Nb and N within the electrodes, Al localized to the barrier and substrate---is chemically consistent with the expected NbN/AlN/NbN heterostructure on $\mathrm{Al_2O_3}$, and growth conditions and source purity changes will be able to reduce them in the future.}


\textcolor{blue}{Does the crystalline barrier JJ performance scale with area and tunnel barrier thickness?  In Fig. \ref{fig:area_scaling} (a) we show the $I$--$V$ characteristics of circular JJs of lithographically defined diameters ranging from 3 to 10~$\mu$m. All junctions retain the characteristic nonlinear $I$--$V$ response, with a zero-bias supercurrent branch and a well-defined quasiparticle gap. The critical current density $J_{\mathrm{c}}= I_{\mathrm{c}}/A =4I_{\mathrm{c}}/(\pi d^{2})$ where $d$ is the lithographic junction diameter, has an average value of $17.66 \pm 4.18~\mathrm{A. cm^{-2}}$ across the different junction sizes. Calculating the $I_{\mathrm{c}}$ from the knee-current of corresponding curves, we find a $\approx 2\%$ asymmetry between the positive- and negative-bias branches.  In contrast, we find the gap voltage to remain nearly independent of junction size, with an average value of $V_{\mathrm{g}}=3.71 \pm 0.06~\mathrm{mV}$, indicating good reproducibility. We experimentally obtain a characteristic voltage $I_{\mathrm{c}}R_{\mathrm{N}} = 2.87 \pm 0.45~\mathrm{mV}$, in good agreement with the theoretical Ambegaokar--Baratoff value $I_{\mathrm{c}}R_{\mathrm{N}} = (\pi/4)V_{\mathrm{g}} \approx 2.91~\mathrm{mV}$ \cite{Ambegaokar1963}. }

\textcolor{blue}{We observe the subgap-to-normal-state resistance ratio falls between  $2 \leq R_{\mathrm{sg}}/R_{\mathrm{N}} \leq 10$ in our epiJJs of 4 - 10 $\mu$m diameters. The smallest junction is far superior: the 3~$\mu$m diameter junction shows $R_{\mathrm{sg}}/R_{\mathrm{N}} \approx 245$, indicating reduced subgap leakage.  The finite spread in $J_{\mathrm{c}}$, $I_{\mathrm{c}}R_{\mathrm{N}}$, and $R_{\mathrm{sg}}/R_{\mathrm{N}}$ among junctions of different dimensions summarized in Table S1 that show a nearly invariant $V_{\mathrm{g}}$ suggests spatial variations in barrier transparency, potentially due to inhomogeneity across the sample.}

\textcolor{blue}{Fig. \ref{fig:area_scaling} (b) shows that the JJ $R_{\mathrm{N}}$ at 300 K and 18 mK scales inversely with the junction area consistent with tunneling through the barrier. The average resistance–area product $RA$ increases 4.74$\times$ from approximately 3.3 $\mathrm{k\Omega .\mu  m^2}$ at 300 K to 15.7 $\mathrm{k\Omega .\mu m^2}$ at 18 mK.  We confirmed that the increase in the $RA$ product upon cooling is not due to the resistivity of the NbN electrodes; it is due to the barrier. We performed temperature-dependent characterization of the resistance and find it shows a tunnel resistance characteristic of Mott variable-range hopping $R_{\rm vrh}(T)=R_{0}\exp[(T_0/T)^{\alpha}]$ with a characteristic energy that corresponds to the oxygen impurity eigenvalue in AlN, and with a concentration of oxygen that corresponds to what we measured in Fig. \ref{fig:TEM} (h) in the AlN tunnel barrier.  We explain this interesting observation in the supplementary section, with the hope that with modified epitaxial conditions and purified sources, the oxygen level can be dramatically reduced and eliminated.}

\textcolor{blue}{In Fig. \ref{fig:area_scaling} (c) we show that the resistance–area product increases exponentially over three orders of magnitude with AlN barrier thickness $t$ with a dependence $RA(t)=RA(0) \times \exp(t/\lambda_{\rm R})$.  Consistent with tunneling, it yields a resistance growth length $\lambda_{\rm R} \approx 1.09~\mathrm{nm}$, with a zero barrier thickness extrapolation $RA(0)=0.0369~\mathrm{k\Omega\,\mu m^2}$.  This corroborates barrier-mediated tunneling in the epitaxial NbN/AlN/NbN junctions.}

\textcolor{blue}{In Fig.~\ref{fig:gap_temperature}(a) we show the temperature-dependent voltage-biased $I-V$ characteristics of a $3~\mu\mathrm{m}$ diameter JJ measured from $20~\mathrm{mK}$ to $8.08~\mathrm{K}$. The temperatures reported here correspond to the fast sample exchange (FSE) probe thermometer reading of the dilution refrigerator.  The reduction of the gap voltage and the increase in subgap quasiparticle current with increasing temperature is consistent with the thermal suppression of superconductivity. The corresponding $dI/dV$ spectra in Fig.~\ref{fig:gap_temperature}(b) shows that as the temperature is increased the dominant conductance peaks associated with the sum-gap voltage progressively move towards zero bias and decrease in amplitude.  We observe several weaker subgap features near $1.85~\mathrm{mV}$ that are pronounced at low temperature and progressively suppressed on warming.  We find that this subgap feature is not because of unequal superconducting gap of the JJ electrodes, but likely due to multiple Andreev reflections with non-uniform tunnel transparency, discussed in the supplementary section.}

\textcolor{blue}{The inset of Fig.~\ref{fig:gap_temperature}(b) shows the superconducting gap extracted from the conductance-peak positions, normalized to its low-temperature value, as a function of reduced temperature. The gap remains nearly constant at low temperature and decreases rapidly on approaching the critical temperature, closely following the independently calculated Bardeen–Cooper–Schrieffer temperature dependence shown by the dash-dotted curve \cite{Bardeen1957}, $ \Delta(T) = \Delta(0) \tanh\!\left[1.74 \sqrt{(T_{\rm c}/T)-1} \right]$.}

\textcolor{blue}{In addition to the experimental evidence already described, two signatures defining the Josephson effect are hysteretic current in underdamped junctions, and Fraunhofer oscillations of the critial supercurrent with magnetic field perpendicular to the junction.  In Fig.~\ref{fig:gap_temperature_fraunhofer}(a) we show the hysteresis in current-biased sweep at zero magnetic field, and in Fig.~\ref{fig:gap_temperature_fraunhofer}(b) we show the measured Fraunhofer modulation.}

\textcolor{blue}{We measured the junction under current bias and examined the magnetic-field dependence of the switching current.  At ${\bf B}=0$, Fig.~\ref{fig:gap_temperature_fraunhofer}(a) shows hysteretic switching between the superconducting and resistive branches, with a maximum switching current of approximately $I_{\mathrm{c}}(0)\approx 0.16~\mu\mathrm{A}$.  Upon application of a magnetic field, we observe $I_{\mathrm{c}}$ develops a pronounced diffraction pattern comprising a dominant central lobe and progressively suppressed side lobes, as shown in Fig.~\ref{fig:gap_temperature_fraunhofer}(b).  Owing to the circular junction geometry, the magnetic-field modulation is expected to follow the Airy diffraction pattern,
\begin{equation}
I_{\mathrm{c}}(\Phi)=
I_{\mathrm{c}}(0)
\left|
\frac{2J_{1}\!\left(\pi\Phi/\Phi_{0}\right)}
{\pi\Phi/\Phi_{0}}
\right|,
\end{equation}
rather than the sinc dependence characteristic of a rectangular Josephson junction \cite{Tinkham1996}.  Here $\Phi$ is the magnetic flux, and $\Phi_{0}=h/(2e)$ its quantum. The measured switching current follows the expected Bessel-like envelope, exhibiting a central lobe together with multiple higher-order side lobes, confirming Josephson transport across the full junction area. When we fit the central lobe of the magnetic-field modulation, we obtain quite reasonable effective magnetic thickness of \(d_{\mathrm{eff}} \approx 0.14\,\mu\mathrm{m}\) and an effective junction radius of \(R \approx 1.17\,\mu\mathrm{m}\).  The Bessel-like diffraction response establishes spatially distributed Josephson coupling through the epitaxial AlN barrier. We have presented further analysis of the observed behavior, including deviations from ideality in the Supplementary Information.} 

\textcolor{blue}{The hysteretic current-biased $I-V$ and Fraunhofer characteristics, considered together with the voltage-biased dc $I-V$ characteristics of Fig. \ref{fig:overview} (c) and (d), and its temperature-dependence in Fig. \ref{fig:gap_temperature}, provide compelling evidence of the successful realization of the first MBE grown NbN/AlN/NbN Josephson junctions based on transport.  That the tunnel barrier is crystalline is seen in Figs. \ref{fig:TEM} (b, c, d and e), and that the JJ current scales with junction area, and with tunnel barrier thickness in Fig. \ref{fig:area_scaling} further corroborates that epitaxial trilayer SIS Josephson junctions with crystalline barriers is achieved by MBE.}

\textcolor{blue}{While the key hurdle of MBE--columnar growth of thick crystalline superconducting layers--was overcome by making thin superconducting electrodes and a nearly lattice-matched crystalline tunnel barrier, some issues such as undesirable oxygen levels and barrier thickness non-uniformity remain unresolved.  We were able to correlate the impact of both these problems with the electonic properties of the JJs: through multiple Andreev reflections, and through temperature dependence of the tunneling current through variable range hopping.  These problems will be addressed with modifications to the epitaxial process and device fabrication.}

\textcolor{blue}{Beyond demonstrating coherent Josephson tunneling through a crystalline nitride barrier, these results motivate further investigation of whether epitaxial control over barrier structure and interfaces can suppress two-level-system (TLS) defects and their associated dielectric losses. The versatility of MBE opens opportunities to adapt established semiconductor heterostructure-engineering strategies, including digital alloys, ternary nitride barriers, and compositionally graded heterostructures, to tailor tunnel-barrier heights, transparency, and electronic properties. Together, these advances should enable more equisite control over Josephson junction characteristics while facilitating their integration with an expanded set of nitride-based superconducting, semiconducting, and quantum devices.}

\section*{Methods}
\subsection{Growth}
We grew the NbN films reported in this work by molecular beam epitaxy (MBE) on commercially annealed (0001) Al$_2$O$_3$ substrates using a Veeco GenXplor MBE system. Prior to growth, the substrates were cleaned sequentially in acetone and 2-propanol for 5 min each. The substrate was then transferred to the growth chamber and heated to a thermocouple temperature of 600$^\circ$C. During growth, 6N5 N$_2$ was supplied at 2 sccm and activated using an RF plasma source operated at 300 W. Nb was supplied from a 4N5-purity source using an electron-beam evaporator, with the Nb flux monitored by electron-impact emission spectroscopy (EIES). The active-N-to-Nb flux ratio was maintained at approximately 1.5, resulting in an NbN growth rate of approximately 1 nm/min. The bottom NbN layer was grown for 10 min to obtain a thickness of approximately 10 nm. AlN was subsequently grown at the same substrate temperature with an Al:N flux ratio of approximately 1:4. The top NbN layer was then deposited under the same growth conditions as the bottom NbN layer.

\subsection{Fabrication}
Device fabrication began with definition of the bottom NbN electrode by BCl$_3$-based inductively coupled plasma reactive-ion etching (ICP-RIE), using an ICP power of 300 W, an RIE power of 8.5 W, a BCl$_3$ flow rate of 30 sccm, and a chamber pressure of 5 mTorr, with the substrate maintained at 10 $^\circ$C. The junction mesa was subsequently defined using the same etch chemistry, with the etch depth controlled to terminate at the bottom NbN electrode. An insulating SiO$_2$ scaffold was then deposited by electron-beam evaporation to electrically isolate the junction sidewalls. Prior to deposition of the top wiring layer, the exposed NbN surface was Ar-ion milled for 15 min to remove the native oxide and ensure a low-resistance electrical contact. The wiring layer consisted of Ti (3 nm)/Nb (200 nm)/Ti (3 nm)/Pt (30 nm), where Ti served as an adhesion layer and Pt provided a protective capping layer.

\subsection{Measurements}
DC transport measurements were performed in a Bluefors LD250 dilution refrigerator with a fast sample exchange (FSE) stage using a QBlox Quantum Source Measurement (QSM) unit for voltage generation and signal acquisition. All the lines used for transport measurements inside the fridge are equipped with a Basel Precision Instruments Microwave-Filter-Thermalizer (MFT) at the mixing stage with an embedded low pass filter having a 3dB cutoff at 2 MHz, which helps in rejecting high frequency noise and lowering the electron temperature. For current-biased measurements, a high-impedance series resistor was used to convert the applied voltage bias into a current bias. The devices were mounted on a custom-made PCB and contacted using Al wire bonds. Four-probe voltages were amplified using a low-noise differential voltage amplifier before acquisition by the QSM voltage measurement channel, while the device current was measured directly with QSM current measurement channel. Temperature-dependent resistance measurements of the superconducting films were performed using a Physical Property Measurement System (PPMS), while the junction resistance from 400 to 80 K was measured using a probe station cooled with liquid nitrogen to perform the temperature sweep.

Triple-axis $\omega$--$2\theta$ XRD scans across the symmetric $(0001)$ wurtzite reflections were measured using a PANalytical Empyrean X-ray diffractometer equipped with a PIXcel3D detector and a Xe proportional detector. The monochromator consisted of a hybrid two-bounce Ge(220) crystal using Cu K$\alpha_1$ radiation. Atomic force microscopy (AFM) measurements were performed using an Asylum Cypher atomic force microscope in tapping mode over scan areas of $2~\mu\mathrm{m} \times 2~\mu\mathrm{m}$. Cross-sectional STEM specimens were prepared using a Thermo Fisher Helios G4 UX focused ion beam (FIB), with a final milling step at 5~keV. STEM measurements were performed using an aberration-corrected Thermo Fisher Spectra 300 CFEG operated at an accelerating voltage of 300~keV.

Synchrotron X-ray reciprocal space mapping (RSM) of the NbN/AlN (5 nm)/NbN/c-Al$_2$O$_3$ film was performed at the Stanford Synchrotron Radiation Lightsource (SSRL), beamline BL 17-2. The map was collected in grazing-incidence geometry with an incidence angle of $\alpha = 0.180^\circ$ and a photon energy of 17 keV. The sample was mounted on a goniometer in front of an Eiger 4M detector and rotated azimuthally using a $\phi$ motor to collect the RSM up to $Q = 6.5~\text{\AA}^{-1}$. The RSM was constructed by binning detector pixels into reciprocal-space intervals of $0.02~\text{\AA}^{-1}$.

\section*{Acknowledgments}
This work is supported by the AFOSR/LPS program Materials for Quantum Computation (MQC) as part of the EpiQ team monitored by Dr. Ali Sayir of AFOSR and Dr. Jacqueline Cooke and Dr. Erin Cleveland of LPS with the award number FA9550-23-1-0688, and partially by an ONR Grant number N00014-22-1-2633 monitored by Dr. Paul Maki.  This work was performed in part at the Cornell NanoScale Facility, a member of the National Nanotechnology Coordinated Infrastructure (NNCI), which is supported by the National Science Foundation (Grant NNCI-2025233). The authors acknowledge the use of facilities and instrumentation supported by NSF through the Cornell University Materials Research Science and Engineering Center DMR-1719875. and magnetic-field modulation of epitaxial NbN/AlN/NbN Josephson junctions.
The Thermo Fisher Spectra 300 X-CFEG was acquired with support from PARADIM, an NSF MIP (DMR-2039380), and Cornell University. N.A.P. acknowledges support from National Science Foundation Graduate Research Fellowship under grant number DGE2139899.
This work is supported in part by the AFOSR Materials (MQC) program under contract No. FA9550-23-1-0692. Use of the Stanford Synchrotron Radiation Lightsource, SLAC National Accelerator Laboratory, is supported by the U.S. Department of Energy, Office of Science, Office of Basic Energy Sciences under Contract No. DE-AC02-76SF00515.

\section*{Data availability}

The data that support the findings of this study are available from the corresponding author upon reasonable request.



\bibliography{References}


\clearpage
\onecolumngrid

\end{document}


\begin{center}

{\bfseries\large Supplementary Information}\\[7pt]

{\bfseries
All-nitride epitaxial SIS Josephson junctions grown by MBE
}\\[10pt]

{\small

\underline{Manas Verma}$^{*}$,
\underline{Anand Ithepalli},
\underline{Amit Rohan Rajapurohita},
Clara Chung,
David A. Garcia-Wetten,
Naomi Pieczulewski,
Danqing Wang,
John G. Wright,
Archit Shah,
Jimy Encomendero,
Rishabh Singh,
Chandrashekar Savant,
Hangwang Yang,
Anthony Hyatt,
Jaehong Choi,
Thomas Varghese,
Kushagra Aggarwal,
Paul B. Welander,
Apurva Mehta,
David Schuster,
Valla Fatemi,
Gregory D. Fuchs,
Michael C. Hamilton,
Hong X. Tang,
David A. Muller,
Huili Grace Xing,
and Debdeep Jena
}\\[8pt]

{\footnotesize
Corresponding author: mv388@cornell.edu

Underlined authors contributed equally.
}

\end{center}

\vspace{10pt}
\tableofcontents
\clearpage
\FloatBarrier

\begin{center}
    {\Large\bfseries Supplementary Information}

    \vspace{0.5cm}

    {\large\bfseries
    ``All-nitride epitaxial SIS Josephson junctions grown by MBE''
    }

    \vspace{0.3cm}

\end{center}

\setcounter{section}{0}
\renewcommand{\thesection}{S\arabic{section}}

\setcounter{figure}{0}
\renewcommand{\thefigure}{S\arabic{figure}}

\setcounter{table}{0}
\renewcommand{\thetable}{S\arabic{table}}

\setcounter{equation}{0}
\renewcommand{\theequation}{S\arabic{equation}}

\section{Additional structural characterization}

\begin{figure}[htb]
    \includegraphics[width = \textwidth]{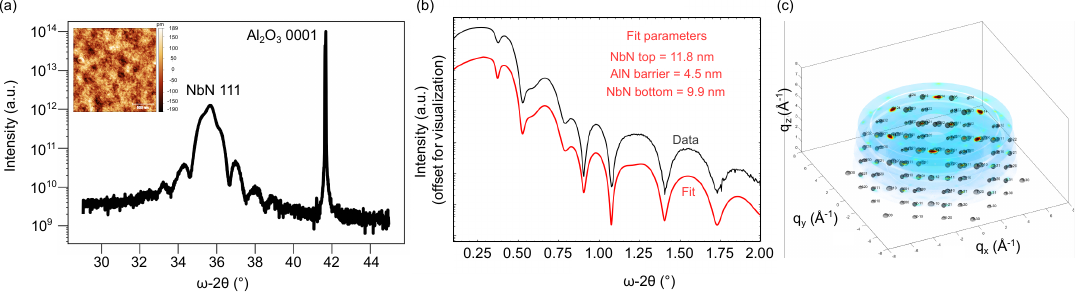}
    \caption{
    \label{fig:S1}
    \textbf{Structural characterization and fabrication of epitaxial NbN/AlN/NbN Josephson junctions.}
\textbf{a}, X-ray diffraction $\omega$--$2\theta$ scan of the NbN/AlN/NbN heterostructure, showing the NbN(111) reflection and the Al$_2$O$_3$(0001) substrate peak. Inset shows a 2 $\mu$m x 2 $\mu$m atomic force microscopy image of the top NbN surface, showing a root-mean-square roughness of 0.06~nm.
\textbf{b}, X-ray reflectivity (XRR)
measurement (black curve) and corresponding multilayer model fit (red curve). 
\textbf{c}, Reciprocal lattice points of the wurtzite AlN crystal structure are plotted and indexed, with forbidden reflections omitted. Slices of the X-ray RSM are overlaid, excluding reflections not attributed to the NbN films. All observed peaks can be attributed to the wurtzite AlN phase.
}
\end{figure}

\begin{figure}[htb]
    \includegraphics[width=\textwidth]{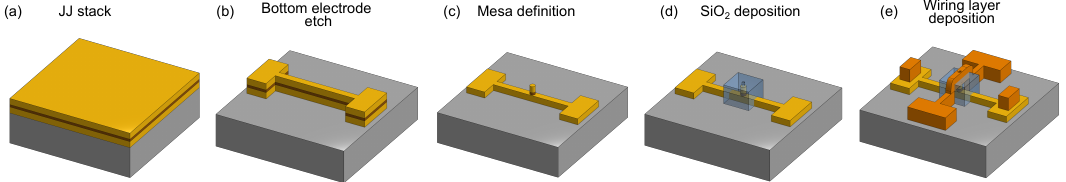}
    \caption{
    \label{fig:S2}
    \textbf{Structural characterization and fabrication of epitaxial NbN/AlN/NbN Josephson junctions.}
\textbf{d--h}, Schematic fabrication sequence, beginning with the as-grown NbN/AlN/NbN trilayer, followed by bottom-electrode etching, junction-mesa definition, SiO$_2$ deposition, and final wiring-layer deposition. 
    }
\end{figure}

Fig. \ref{fig:S1} (a) shows the $\omega$--$2\theta$ X-ray diffraction scan. The expected Al$_2$O$_3$ substrate reflections together with the NbN(111) reflection, confirms the out-of-plane orientation of the NbN electrodes. Pronounced finite-thickness fringes are observed around the NbN reflection, indicating a high degree of crystalline coherence and well-defined interfaces within the heterostructure. The atomic force microscopy (AFM) of the top NbN surface as shown in the inset yields a low root-mean-square roughness of $\leq 0.1$~nm, demonstrating that the smooth morphology of the heterostructure is maintained through the complete trilayer growth. Fig. \ref{fig:S1} (b) shows the X-ray reflectivity (XRR) measurement
(black curve) and the corresponding multilayer model fit (red
curve). The well-defined interference oscillations are closely
reproduced by the fitted model, indicating well-defined layer
thicknesses and interfaces. The extracted thicknesses of 11.8
nm, 4.5 nm, and 9.9 nm for the top NbN electrode, AlN barrier,
and bottom NbN electrode, respectively, are in close agreement
with the targeted growth thickness. Fig. \ref{fig:S1} (c) shows the schematic representation of the reciprocal lattice points for the wurtzite AlN and cubic NbN crystal structures, with symmetry-forbidden reflections omitted. Selected slices of the experimentally measured X-ray reciprocal space map (RSM) are overlaid to identify the observed diffraction features. All additional reflections not indexed to the cubic NbN lattice can be consistently attributed to the wurtzite AlN phase, confirming the presence of the wurtzite structure within the NbN/AlN/NbN heterostructure.

Fig. \ref{fig:S2} (a)--(e) illustrates the device fabrication process. Starting from the as-grown NbN/AlN/NbN trilayer, the bottom electrode was patterned and etched using BCl$_3$-based inductively coupled plasma reactive-ion etching (ICP-RIE), with a BCl$_3$ flow rate of 30 sccm, chamber pressure of 5 mTorr, ICP power of 300 W, and RIE power of 8.5 W, resulting in a DC self-bias of approximately 78 V. Backside He cooling was employed to maintain the substrate chuck at 10 $^\circ$C during etching. The junction mesa was subsequently defined using the same etching conditions, with a carefully controlled etch duration to terminate within the bottom NbN electrode. As the etching process was nonselective between the constituent layers, precise timing was required to preserve the bottom electrode while defining the active junction area. A 130-nm-thick SiO$_2$ insulating layer was then selectively deposited around the junction mesa by electron-beam evaporation to electrically isolate the junction sidewalls from the subsequent wiring layer. Prior to wiring layer deposition, a 15-min Ar-ion sputter cleaning step was performed to remove native surface oxides and ensure electrical contact to the NbN electrodes. Finally, a Ti (3 nm)/Nb (200 nm)/Ti (3 nm)/Pt (30 nm) wiring stack was deposited by DC sputtering and patterned to establish electrical connections to the top and bottom NbN electrodes, yielding individually addressable NbN/AlN/NbN Josephson junctions.

\section{Additional electrical characterization}

\begin{figure}[!htb]
    \includegraphics[width=\textwidth]{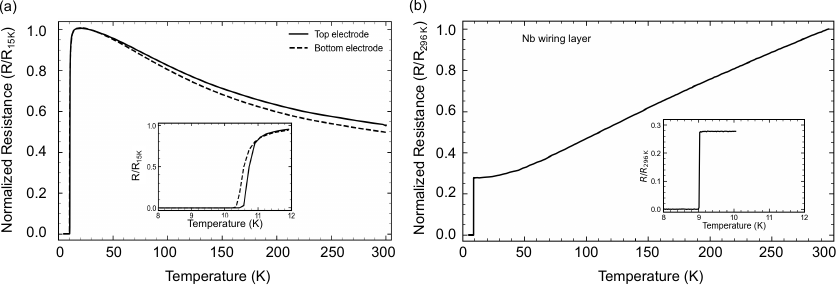}
    \caption{
    \label{fig:S3}
   \textbf{Temperature-dependent resistance of the NbN electrodes and wiring layer.}
\textbf{a}, Normalized resistance, $R/R_{15\,\mathrm{K}}$, as a function of temperature for the top (solid) and bottom (dashed) NbN electrodes of the NbN/AlN/NbN heterostructure. Both electrodes exhibit metallic normal-state behavior and superconducting transitions near 11~K. The inset highlights the superconducting transitions of the two electrodes.
\textbf{b}, Normalized resistance, $R/R_{300\,\mathrm{K}}$, of the Nb wiring layer as a function of temperature, showing metallic behavior followed by a superconducting transition near 9~K. The inset shows an expanded view of the superconducting transition.
    }
\end{figure}

Fig. \ref{fig:S3} shows the temperature-dependent electrical transport of the individual NbN electrodes and the superconducting wiring layer used in the junction devices. Fig. \ref{fig:S3} (a) presents the normalized resistance of the top and bottom NbN electrodes as a function of temperature. Both electrodes exhibit the resistance increasing upon cooling, followed by superconducting transitions near 11 K. The closely spaced transition temperatures indicate that both NbN layers retain comparable superconducting properties despite their different positions within the NbN/AlN/NbN heterostructure. The inset highlights the superconducting transitions and the small difference in $T_{\mathrm{c}}$ between the two electrodes. Fig. \ref{fig:S3} (b) shows the corresponding temperature-dependent resistance of the Nb wiring layer used for electrical contact to the junctions. The Nb layer exhibits metallic normal-state transport and undergoes a superconducting transition near 9 K, as highlighted in the inset. These measurements confirm that both epitaxial NbN electrodes and the Nb wiring layer are superconducting at the temperatures used for the low-temperature Josephson-junction measurements.

\begin{figure}
    \includegraphics[width=\textwidth]{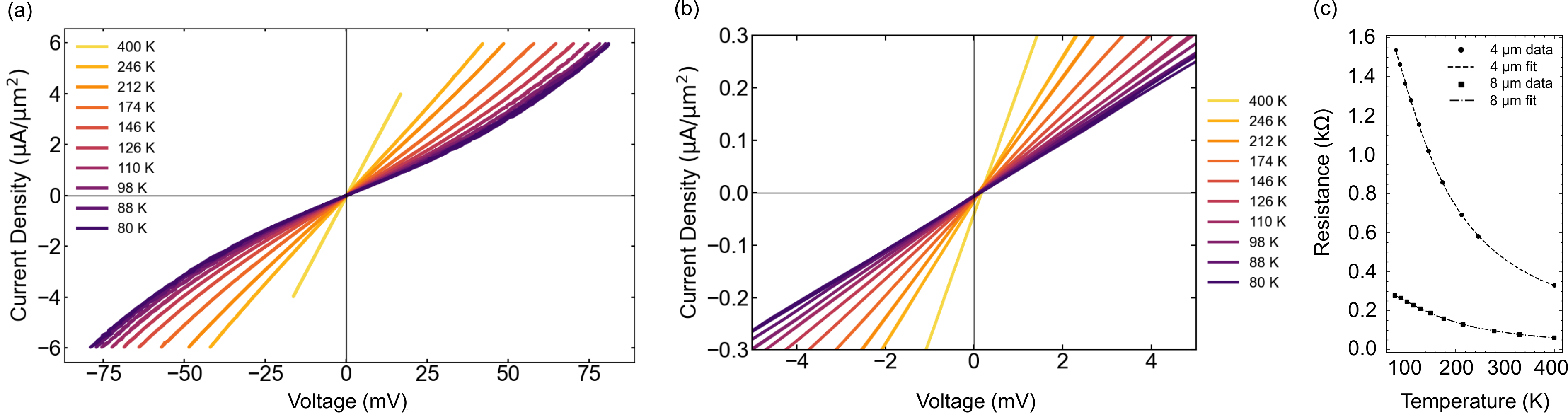}
    \caption{
    \label{fig:S4}
   \textbf{Temperature-dependent current--voltage characteristics of a 4-\textmu m-diameter NbN/AlN/NbN junction.}
\textbf{a}, Current-density--voltage ($J$--$V$) characteristics measured in a probe station from 80 to 400~K, showing a systematic increase in junction resistance with decreasing temperature.
\textbf{b}, Enlarged view of the low-bias region, which remains approximately linear between $-5$ and $+5$~mV over the measured temperature range. The temperature-dependent junction resistance was extracted by linear fitting of the $I$--$V$ data within this low-bias interval and used for the subsequent transport analysis. The small offset of the $I$--$V$ characteristics from the origin is a measurement artifact and does not reflect an intrinsic junction response.
    }
\end{figure}

Fig. \ref{fig:S4} shows the temperature-dependent electrical transport of a 4-\textmu m-diameter NbN/AlN/NbN junction measured in a probe station over the temperature range of 80--400~K. The $J$--$V$ characteristics [Fig. \ref{fig:S4} (a)] exhibit a systematic reduction in current at a given bias upon cooling, corresponding to an increase in the junction resistance with decreasing temperature. To quantitatively determine this temperature dependence, the low-bias region of the $I$--$V$ characteristics is shown in Fig. \ref{fig:S4} (b). Over the bias range from $-5$ to $+5$~mV, the response remains approximately linear at all measured temperatures, allowing the junction resistance to be extracted from linear fits to this region. The resulting resistance values are used for the temperature-dependent transport analysis. A small displacement of the measured $I$--$V$ curves from the origin is attributed to an instrumental measurement offset and does not represent an intrinsic electrical response of the junction. 
\\

\section{Multiple Andreev Reflections}

An interpretation of the $\sim 1.85$ mV subgap feature observed in Fig. 5 in terms of unequal superconducting gaps is not supported by the independently measured transition temperatures. As shown in Fig. S3 (a), the pristine bottom and top NbN electrodes exhibit $T_{\mathrm{c}} = 10.48$ K and $10.82$ K, respectively, while the Nb/Ti wiring has $T_{\mathrm{c}} \sim 9$ K (Fig. S3 (b)), as determined from PPMS measurements. To account for both the $\sim 3.7$ mV sum-gap voltage and the $\sim 1.85$ mV feature through a difference-gap mechanism, the superconducting gaps would need to be approximately $\Delta_1 = 2.775$ meV and $\Delta_2 = 0.925$ meV, corresponding to a gap ratio of $\sim 3$. However, the transition temperatures of the two NbN electrodes differ by only $\sim 3\%$. Even when the Nb/Ti wiring is considered, the measured transition temperatures are too similar to explain a gap ratio of $\sim 3$, assuming comparable gap-to-$T_{\mathrm{c}}$ ratios.

These observations make a difference-gap interpretation unlikely. Instead, the observed subgap features are consistent with subharmonic gap structures arising from multiple Andreev reflections (MAR) corresponding to \(2\Delta/2e\). Further supporting this interpretation, an additional weaker structure is observed at \(V \approx 1.23\) mV, close to the expected \(n=3\) subharmonic at \(2\Delta/3e\). As discussed by Ternes et al. \cite{Ternes2006}, for superconducting electrodes with nearly equal energy gaps, MAR-related structures are expected at (2$\Delta$/e), (2$\Delta$/2e), and (2$\Delta$/3e) for the first three orders. Similar subharmonic features have previously been reported in NbN/AlN/NbN \cite{Wang1994, Wang1997}, NbN/MgO/NbN \cite{Uzawa2005, Uzawa2007}, and Nb/AlO\(_x\)/Nb junctions \cite{Kleinsasser1993, Kleinsasser1994} and attributed to MAR processes. In nominally tunnel-like junctions, such MAR features have been linked to barrier inhomogeneity, where small locally thinner regions provide higher-transparency channels that support Andreev transport.  For a homogeneous tunnel barrier with a single-particle transparency $\mathcal{T}$, an $n$th-order MAR process involves $n$ traversals of the barrier, such that the corresponding current step scales approximately as \cite{Octavio1983, Blonder1982, Uzawa2007}
\begin{equation}
J_\mathrm{n} \propto \mathcal{T}^{\,\mathrm{n}}.
\end{equation}
Consequently, the ratio between successive MAR current steps is expected to be
\begin{equation}
\frac{J_\mathrm{n}}{J_\mathrm{n-1}} \propto \mathcal{T},
\end{equation}
and therefore, for a spatially homogeneous barrier \cite{vanderPost1994, Bratus1995},
\begin{equation}
\frac{J_2}{J_1} \simeq \frac{J_3}{J_2}.
\end{equation}
In our junction, however, the extracted ratios are $J_2/J_1=5.33$ and $J_3/J_2=3.3$, which differ substantially. This departure from the expected MAR scaling suggests a non-uniform tunnel transparency, consistent with an inhomogeneous AlN barrier containing localized higher-transparency conduction regions \cite{Uzawa2007}.

\begin{figure}[htbp]
    \centering
    \includegraphics[width=\textwidth]{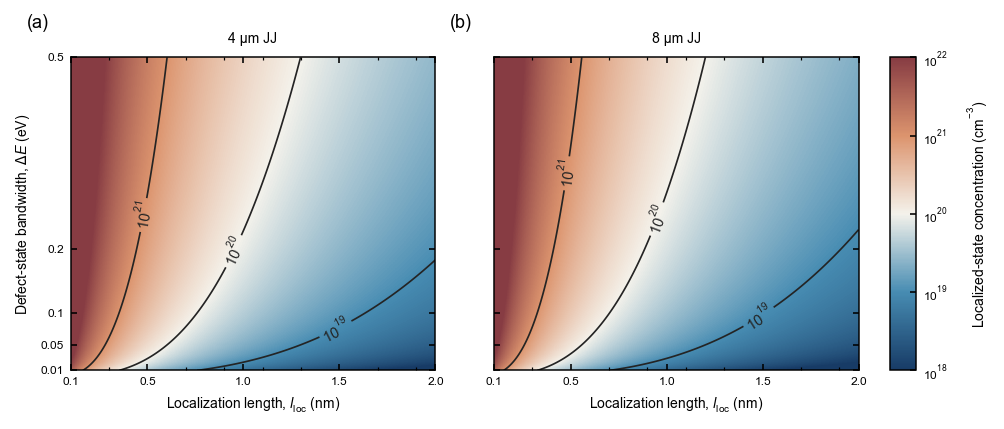}
    \caption{\textbf{Estimated localized-state concentration from Mott variable-range hopping analysis.} Calculated localized-state concentration, \(g(E_{\mathrm{F}})\Delta E\), as a function of localization length, \(l_{loc}\), and defect-state bandwidth, \(\Delta E\), for \textbf{a}, 4 $\mu$m and \textbf{b}, 8 $\mu$m Josephson junctions. The estimates are based on the characteristic Mott temperatures extracted from temperature-dependent resistance measurements. Solid black contours indicate localized-state concentrations of \(10^{19}\), \(10^{20}\), and \(10^{21}\,\mathrm{cm^{-3}}\). The estimated concentrations are comparable to the oxygen and hydrogen concentrations measured by SIMS over plausible parameter ranges, suggesting a possible contribution from oxygen or hydrogen-related localized states to hopping-assisted transport.}
    \label{fig:S5}
\end{figure}

\section{Estimation of localized-state concentration from Mott variable-range hopping analysis}

Fig. S4 (c) shows the temperature dependence of the normal-state resistance investigated for 4 $\mu$m and 8 $\mu$m-diameter NbN/AlN/NbN junctions measured across a temperature range from 80 K to 400 K. All devices characterized in this work were fabricated on the same chip, except those measured for this plot (Fig. S4 (c)), which were fabricated on a separate chip. Both chips were co-fabricated in the same processing run from the same MBE-grown wafer. At each temperature, the junction resistance was obtained from a linear fit to the $I$-$V$ characteristic between $-5$ and $+5$~mV, over which the response was approximately linear as shown in Fig. S4 (b). Both devices exhibit an increase in resistance upon cooling, indicating the presence of a thermally activated contribution to transport in addition to the weakly temperature-dependent direct tunneling conductance of the junction.

To describe this behavior, the resistance was modeled using two parallel conduction channels: a temperature-independent junction resistance,
$R_{\mathrm{TI}}$, and a temperature-dependent variable-range-hopping (VRH) resistance \cite{Mott1969, Efros1975, Ruhtinas2025},
\begin{equation}
R_{\mathrm{VRH}}(T)
=
R_{0}
\exp\left[
\left(\frac{T_{0}}{T}\right)^{\alpha}
\right],
\end{equation}
where $R_{0}$, $T_{0}$, and $\alpha$ were treated as free fitting parameters. The total resistance is therefore
\begin{equation}
\frac{1}{R(T)}
=
\frac{1}{R_{\mathrm{TI}}}
+
\frac{1}{R_{\mathrm{VRH}}(T)},
\end{equation}

The model plotted along with the measured data in Fig. S4 (c) reproduces the measured $R(T)$ characteristics well for both junction sizes. The extracted hopping exponents are $\alpha=0.309$ and
$\alpha=0.277$ for the 4 $\mu$m and 8 $\mu$m junctions respectively, close to the $\alpha=1/4$ dependence expected for three-dimensional Mott variable-range hopping \cite{Mott1969}. This behavior is consistent with an additional conduction channel mediated by hopping between localized electronic states within the AlN tunnel barrier. 

The fitted temperature-independent resistances are approximately $R_{\mathrm{TI}}=1.9$~k$\Omega$ and $330~\Omega$ for the 4 $\mu$m and
8 $\mu$m junctions, respectively. These values may be compared with the normal-state resistances (above $V_{\mathrm{gap}}$ of 3.7 mV) independently measured in the dilution refrigerator, $R_{\mathrm{N}}=1164~\Omega$ and $337~\Omega$, respectively. The consistency of these independent measurements, particularly the close agreement for the 8 $\mu\mathrm{m}$ junction, support the identification of the temperature-independent channel with direct tunneling through the junction barrier. Including the finite-temperature Simmons correction for the tunneling channel produces only a minor change in the fitted \(R(T)\) behavior. The extracted VRH parameters remain essentially unchanged, indicating that the dominant temperature dependence arises from the hopping channel rather than from the intrinsic temperature dependence of direct tunneling. Together, these measurements demonstrate reproducible area scaling, enabling the junction resistance and current to be systematically controlled through the lithographically defined device area.

The thermally activated hopping contribution yields characteristic Mott temperatures of \(T_0 \approx 4.8 \times 10^5\) K and \(6.1 \times 10^5\) K, respectively for 4 $\mu$m and 8 $\mu$m-diameter devices. These parameters are used here to estimate the concentration of localized electronic states and examine their possible association with defects within the AlN barrier. Within the Mott VRH framework, the characteristic temperature is related to the density of localized states near the Fermi level through \cite{Mott1969}

\begin{equation}
T_0 = \frac{18.1}{k_{\mathrm{B}}g(E_{\mathrm{F}})(l_{\rm loc})^3},
\label{eq:mott_T0}
\end{equation}

\noindent where $g(E_{\mathrm{F}})$ is the density of localized states per unit energy and volume evaluated at the Fermi level, \(E_{\mathrm F}\), around which hopping between localized states occurs,, and $l_{\rm loc}$ is the localization length. Assuming a characteristic defect-state energy bandwidth $\Delta E$, the corresponding concentration of localized states can be estimated as

\begin{equation}
N_{\mathrm{loc}} = g(E_{\mathrm{F}})\Delta E
= \frac{18.1\,\Delta E}
{k_{\mathrm{B}}T_0(l_{\rm loc})^3}.
\label{eq:localized_states}
\end{equation}

\noindent
Fig. S5 (a) and (b) show the estimated localized-state concentration, \(g(E_{\mathrm F})\Delta E\), as a function of localization length, \(l_{\rm loc}\), and defect-state energy bandwidth, \(\Delta E\), for the 4 and 8 $\mu$m Josephson junctions, respectively. The solid black contours represent localized-state concentrations of \(10^{19}\), \(10^{20}\), and \(10^{21}\,\mathrm{cm^{-3}}\), illustrating the dependence of the estimated concentration on the assumed localization length and energy bandwidth. For a representative localization length of \(l_{\rm loc} \approx 0.5\) nm and \(\Delta E \approx 30\)–\(36\) meV, the estimated concentrations are of the order of \(10^{20}\,\mathrm{cm^{-3}}\), comparable to the oxygen and hydrogen concentrations measured by secondary-ion mass spectrometry (SIMS) in Fig. 3 (e). This order-of-magnitude agreement suggests that oxygen- and hydrogen-related defects may contribute to the localized electronic states involved in hopping-assisted transport. However, the microscopic origin of these states remains unresolved, and contributions from other point defects, interface disorder, or deviations from ideal AlN stoichiometry cannot be excluded.

\section{Effect of Ar-ion milling and thicker AlN barrier}

\begin{figure}
    \centering
    \includegraphics[width=\textwidth]{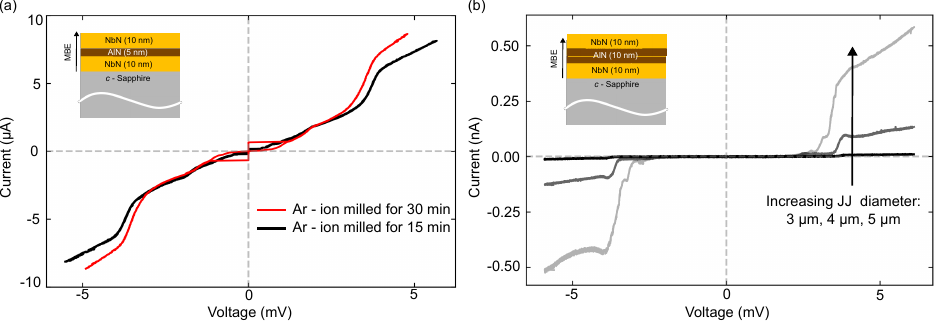}
    \caption{\textbf{Control measurements of NbN/AlN/NbN tunnel junctions.}
    \textbf{(a) 5 nm AlN}: $I$--$V$ characteristics of 5 - \textmu m diameter junctions for which the top NbN electrode was Ar-ion milled for 15~min and 30~min prior to wiring-layer deposition. The nearly overlapping characteristics demonstrate that the measured junction response is insensitive to the milling duration, supporting its origin in the buried AlN barrier rather than the top-electrode contact interface. The red and black curves correspond to current-biased and voltage-biased measurements, respectively.
    \textbf{(b) 10 nm AlN}: 
    Voltage-biased $I$--$V$ characteristics of junctions incorporating a 10-nm-thick AlN barrier for junction diameters of 3, 4, and 5~$\mu$m. The devices retain SIS-like characteristics, with increasing resistance for decreasing junction area and an approximately invariant characteristic gap voltage. Insets show schematics of the corresponding NbN/AlN/NbN trilayer structures.}
    \label{fig:S6}
\end{figure}

To verify that the measured junction characteristics originate from transport across the AlN barrier rather than from the top-electrode contacting process, we performed control measurements with different \textit{in-situ} Ar-ion milling durations prior to deposition of the wiring layer. For the devices discussed in the main text, the top NbN electrode was Ar-ion milled for 15 min to remove the native surface oxide, while an additional device was subjected to an extended 30 min milling step. As shown in Fig. \ref{fig:S6} (a), the resulting \(I\)--\(V\) characteristics from a are nearly overlapping and exhibit the same nonlinear behavior for a 5-\textmu m-diameter device, demonstrating that the junction response is insensitive to the milling duration and supporting its origin in the buried AlN tunnel barrier rather than the NbN/wiring-layer interface. The red and black curves correspond to current-biased and voltage-biased measurements, respectively. As an additional control, junctions with a thicker, 10-nm AlN barrier were fabricated as shown in Fig. \ref{fig:S6} (b). These devices retain characteristic SIS-like nonlinear \(I\)--\(V\) behavior, with the normal-state resistance increasing systematically with decreasing junction area, while the characteristic gap voltage remains approximately unchanged across the different junction diameters. Together, these measurements support tunnelling through the AlN barrier as the dominant origin of the observed junction characteristics.

\section{Effect of thin electrodes on the magnetic-field dependence of critical current}
The magnetic-field dependence of the critical current provides insight into the spatial distribution of supercurrent across a Josephson junction. For an ideal junction with a uniform critical current density and negligible self-field effects, the critical current follows the conventional Fraunhofer interference pattern. Although the NbN/AlN/NbN junctions presented in this work exhibit clear magnetic-field-dependent modulation of the critical current, the observed interference pattern deviates from the ideal Fraunhofer response. Here, these deviations are examined in greater detail, with particular emphasis on the possible influence of nonuniform current distribution and the thin-film geometry of the superconducting electrodes.

Using the normal-state resistivity $\rho \approx 150~\mu\Omega\,\mathrm{cm}$ and critical temperature $T_c = 10.8~\mathrm{K}$, the dirty-limit magnetic penetration depth \cite{Zmuidzinas2012, Ruhtinas2025, Mattis1958} is estimated as
\[
\lambda \approx 105~\mathrm{nm}
\sqrt{\frac{\rho~(\mu\Omega\,\mathrm{cm})}{T_\mathrm{c}~(\mathrm{K})}}
\approx 391~\mathrm{nm},
\]
similar to values obtained in \cite{Oates1991}. For NbN electrode thicknesses $t_{\rm b} = 9.9~\mathrm{nm}$ and $t_{\rm t} = 11.8~\mathrm{nm}$, the ratios $t/\lambda$ are
\[
\frac{t_{\rm b}}{\lambda}\approx0.025,
\qquad
\frac{t_{\rm t}}{\lambda}\approx0.030,
\]
confirming that both electrodes are deep in the thin-film limit, $t\ll\lambda$. The corresponding Pearl lengths \cite{Pearl1964},
\[
\Lambda=\frac{2\lambda^2}{t},
\]
are $\Lambda_{\rm b}\approx31~\mu\mathrm{m}$ and $\Lambda_{\rm t}\approx26~\mu\mathrm{m}$. Hence, for the $D=3~\mu\mathrm{m}$ junction,
\[
\frac{D}{\Lambda_\mathrm{b}}\approx0.10,
\qquad
\frac{D}{\Lambda_\mathrm{t}}\approx0.12,
\]
indicating that the junction dimensions are substantially smaller than the thin-film screening length, so nonlocal magnetic screening and field redistribution may influence the diffraction pattern \cite{Rosenthal1991, Boris2013, Moshe2008, Clem2010, Ruhtinas2025, Mints1995}. In addition, even a small out-of-plane magnetic-field component, arising for example from a slight misalignment between the applied field and the junction plane, can modify the magnetic interference pattern \cite{Mawatari2022, Monaco2013}. For the thin NbN electrodes, the magnetic thickness relevant to Josephson self-fields is \cite{Monaco2013}
\[
d_\mathrm{j}=d_\mathrm{I}+\lambda_\mathrm{b}\coth\!\left(\frac{t_\mathrm{b}}{\lambda_\mathrm{b}}\right)
+\lambda_\mathrm{t}\coth\!\left(\frac{t_\mathrm{t}}{\lambda_\mathrm{t}}\right),
\]
which, using $d_{\rm I}=4.5~\mathrm{nm}$ and $\lambda_{\rm b}\simeq\lambda_{\rm t}\simeq391~\mathrm{nm}$, gives $d_j\approx28.4~\mu\mathrm{m}$. The Josephson penetration depth is therefore 
\[
\lambda_\mathrm{J}=
\sqrt{\frac{\Phi_0}{2\pi\mu_0 J_\mathrm{c} d_\mathrm{j}}}
\approx8.41~\mu\mathrm{m},
\]
for $J_\mathrm{c}\approx13~\mathrm{A\,cm^{-2}}$. Thus,
\[
\frac{D}{\lambda_\mathrm{J}}\approx0.36<1,
\]
placing the $3~\mu\mathrm{m}$ device on the small-junction side of the self-field crossover. Josephson self-fields are therefore expected to provide only a moderate correction, whereas the much stronger inequalities $t\ll\lambda$ and $D\ll\Lambda$ indicate that thin-film electromagnetic effects, together with possible spatial nonuniformity of the current or magnetic-field distribution, are more likely to account for deviations from the ideal Airy diffraction response \cite{Rosenthal1991, Boris2013, Moshe2008, Clem2010, Ghatak2018, Ruhtinas2025, Mints1995, Owen1967}.

\begin{table}[p]
\caption{\textbf{Electrical parameters of NbN / AlN (5 nm) / NbN Josephson junctions measured at 18 mK.}}
\label{tab:table_S1}
\centering
\footnotesize
\setlength{\tabcolsep}{4.5pt}
\renewcommand{\arraystretch}{1.15}

\begin{adjustbox}{max width=\textwidth}
\begin{tabular}{ccccccccccccccc}
\toprule
\makecell{Junction} &
\makecell{$d$ \\ ($\mu$m)} &
\makecell{$J_c^{+}$ \\ (A cm$^{-2}$)} &
\makecell{$J_c^{-}$ \\ (A cm$^{-2}$)} &
\makecell{$V_g^{+}$ \\ (mV)} &
\makecell{$V_g^{-}$ \\ (mV)} &
\makecell{$V_m^{+}$ \\ (mV)} &
\makecell{$V_m^{-}$ \\ (mV)} &
\makecell{$R_N$ \\ ($\Omega$)} &
\makecell{$R_{\mathrm{sg}}$ \\ ($\Omega$)} &
\makecell{$R_{\mathrm{sg}}/R_N$} &
\makecell{$I_cR_N^{+}$ \\ (mV)} &
\makecell{$I_cR_N^{-}$ \\ (mV)} &
\makecell{$I_cR_N^{\mathrm{th},+}$ \\ (mV)} &
\makecell{$I_cR_N^{\mathrm{th},-}$ \\ (mV)} \\
\midrule

1 & 3  & 12.87 & $-13.08$ & 3.726 & $-3.696$ & 576.81 & $-586.24$ & 2587.47 & 634055.96 & 245.05 & 2.354 & $-2.392$ & 2.926 & $-2.902$ \\

2 & 4  & 23.12 & $-24.31$ & 3.782 & $-3.859$ & 9.026 & $-9.490$ & 1164.05 & 3106.65 & 2.669 & 3.382 & $-3.556$ & 2.970 & $-3.031$ \\

3 & 5  & 20.38 & $-19.31$ & 3.757 & $-3.686$ & 9.358 & $-8.867$ & 799.55 & 2338.71 & 2.925 & 3.199 & $-3.031$ & 2.951 & $-2.895$ \\

4 & 8  & 14.52 & $-14.82$ & 3.646 & $-3.678$ & 19.375 & $-19.785$ & 337.21 & 2655.64 & 7.875 & 2.460 & $-2.512$ & 2.864 & $-2.889$ \\

5 & 10 & 17.44 & $-18.89$ & 3.668 & $-3.703$ & 6.202 & $-6.720$ & 215.94 & 452.89 & 2.097 & 2.957 & $-3.204$ & 2.881 & $-2.908$ \\

\bottomrule
\end{tabular}
\end{adjustbox}

\end{table}


\section{Measurement schematic}

\begin{figure}
    \centering
    \includegraphics[width=\linewidth]{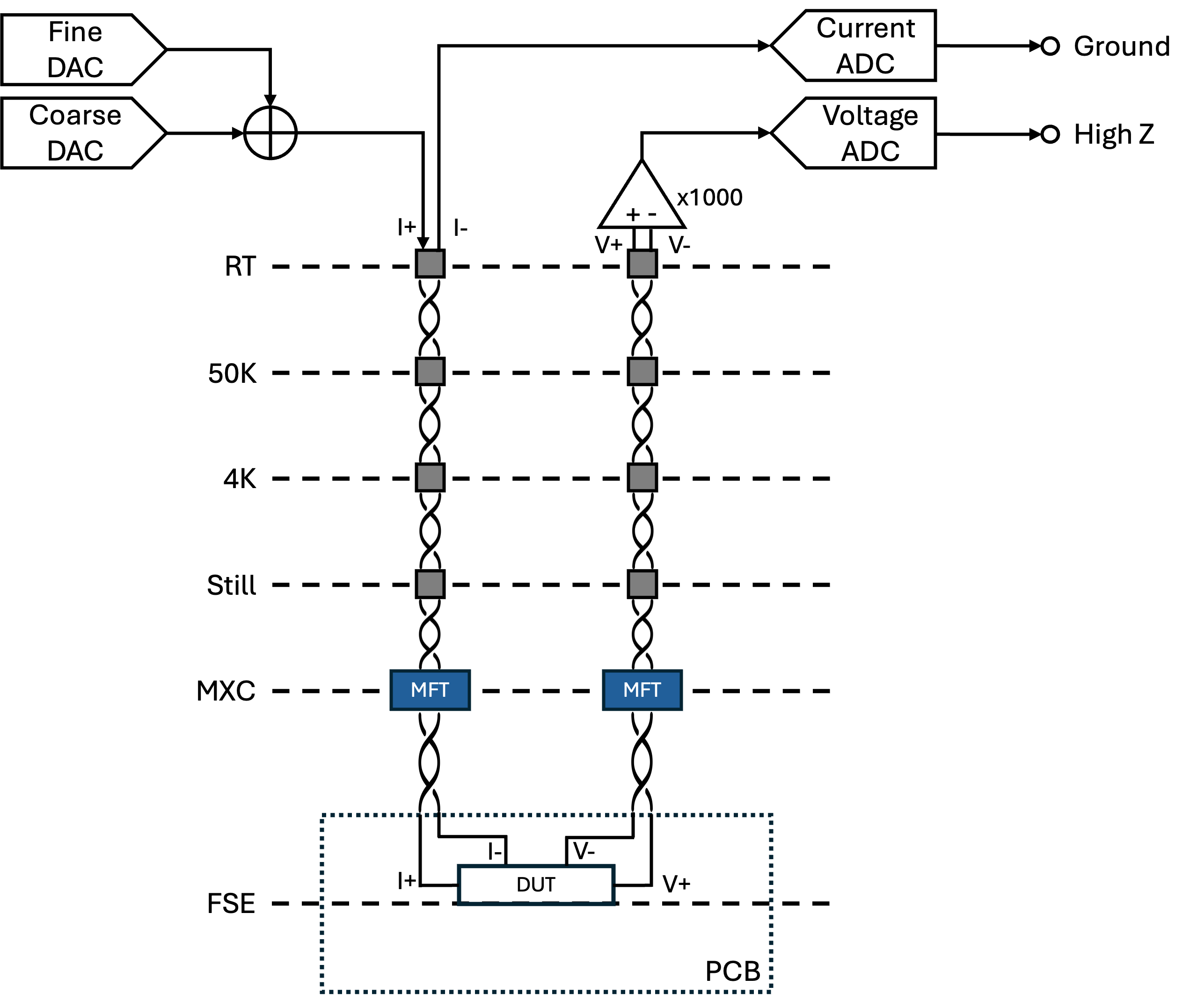}
    \caption{\textbf{DC measurement schematic inside the dilution refrigerator (DR):} DC transport measurements were performed in a dilution refrigerator (DR). The device bias was generated by summing coarse and fine digital-to-analog converter (DAC) outputs from a Qblox Quantum Source Measurement (QSM) unit. For current-biased measurements, a high-value series resistor was placed before the DR. Twisted-pair wiring was thermally anchored at each temperature stage using bobbins to improve electron thermalization. At the mixing chamber (MXC) stage, the measurement lines were additionally filtered and thermalized using a Basel Precision Instruments microwave filter and thermalizer (MFT), with an in-line resistance of approximately $30~\mathrm{m}\Omega$ and a 3-dB low-pass cutoff frequency of 2 MHz. The signals were routed to the fabricated Josephson junctions through a custom printed circuit board (PCB) sample holder with aluminium wire-bond connections. The four-terminal voltage signal was preamplified using a Basel Precision Instruments low-noise, low-drift differential voltage amplifier before being acquired by the QSM voltage ADC. The current through the device was measured from the $I^{-}$ lead using the QSM current ADC.
}
    \label{fig:S7}
\end{figure}

\clearpage
\printbibliography[title={References}]